\documentclass[aps,pra,superscriptaddress,twocolumn,a4paper,reprint,eprint,longbibliography,footinbib,floatfix]{revtex4-1}
\pdfoutput=1
\usepackage[percent]{overpic}
\usepackage{physics}
\usepackage[utf8]{inputenc}
\usepackage{graphicx}
\usepackage{amsmath}
\usepackage{amsfonts}
\usepackage{amssymb}
\usepackage{bbm}
\usepackage{bm}
\usepackage{bbold}
\usepackage[usenames,dvipsnames,svgnames,table]{xcolor}
\usepackage[USenglish]{babel}
\usepackage{hyperref}
\usepackage{grffile}
\usepackage{multirow}
\usepackage{hhline}
\usepackage[normalem]{ulem}

\newcommand{\ignore}[1]{}

\allowdisplaybreaks

\pdfpageattr{/Group <</S /Transparency /I true /CS /DeviceRGB>>}

\begin{document}

\title{Bardasis-Schrieffer-like phase mode in a superconducting bilayer} 

\author{Nico A. Hackner}
\author{P. M. R. Brydon}
\email{philip.brydon@otago.ac.nz}
\affiliation{Department of Physics and MacDiarmid Institute for
Advanced Materials and Nanotechnology, University of Otago, P.O. Box
56, Dunedin 9054, New Zealand}

\date{July 4, 2023}

\begin{abstract}
We theoretically study the low-lying collective modes of an even-parity spin-singlet superconducting bilayer, where strong spin-orbit coupling leads to a closely competing odd-parity pairing state. We develop a gauge-invariant theory for the coupling of phase fluctuations to an external electromagnetic field and show that the competing odd-parity pairing instability gives rise to a Bardasis-Schrieffer-like phase mode within the excitation gap. Accounting for the long-range Coulomb interaction, however, we find that this mode is converted into an antisymmetric plasmon and is likely pushed into the quasiparticle continuum.
\end{abstract}
\maketitle

{\it Introduction.} The first-order field-induced transition within the superconducting state of CeRh$_2$As$_2$~\cite{khim2021,Landaeta2022} has been interpreted as a transition between even- and odd-parity pairing~\cite{Mockli2021,Schertenleib2021,Cavanagh2022,Nogaki2022}. This requires a near-degeneracy of these different pairing channels, which naturally arises from the sublattice structure of the unit cell~\cite{Fischer2011,Yoshida2012,Yoshida2014}. Specifically, for on-site singlet pairing, even- and odd-parity states can be constructed by setting the pair potential to have the same (``uniform") or opposite (``staggered") sign on the two sublattices, respectively. The staggered state is suppressed by intersublattice hopping, but its critical temperature may nevertheless be comparable to that of the uniform state for sufficiently strong spin-orbit coupling (SOC).

The key evidence for the uniform-staggered transition in CeRh$_2$As$_2$ is its quantitative agreement with the dependence of the phase diagram on the magnitude and orientation of the magnetic field~\cite{khim2021,Landaeta2022}. However, more direct evidence of the staggered state is lacking, and alternative scenarios have been proposed~\cite{Mockli2021,Hazra2022,Machida2022}. Indeed, although sublattice degrees of freedom are considered important for the electronic structure in many superconductors, e.g. Cu$_x$Bi$_2$Se$_3$~\cite{Fu2010}, SrPtAs~\cite{Youn2012}, UPt$_3$\cite{Yanase2016}, bilayer transition metal dichalcogenides (TMDs)~\cite{Nakamura2017,Liu2017}, and UTe$_2$\cite{Shishidou2021}, only CeRh$_2$As$_2$ displays the uniform-staggered transition. Attempts at engineering the uniform-staggered transition in artificial superlattices have also been unsuccessful~\cite{Goh2012}. It is therefore unclear if the staggered state is relevant to the physics of such materials, while the extreme magnetic fields at which it is expected to appear makes studying it a formidable challenge. This motivates us to search for evidence of the staggered state at vanishing magnetic field strength, where the uniform phase is realized.

It has recently been pointed out in Ref.~\cite{Lee2022} that the presence of a subdominant pairing state in CeRh$_2$As$_2$ should give rise to a low-lying Bardasis-Schrieffer (BS) collective mode~\cite{Bardasis1961}, corresponding to fluctuations from the uniform to the staggered state. Moreover, the strong SOC in this material could allow for the optical excitation of this mode, despite the opposite parity of the two pairing states. The first observation of BS modes, which has only recently been claimed~\cite{Bohm2014}, has stimulated much attention~\cite{Maiti2016,Muller2019,Allocca2019,Sun2020,Muller2021,Curtis2022}. The prospect that superconductors with sublattice degrees of freedom generically host BS modes, potentially accessible by terahertz spectroscopy~\cite{Shimano2020}, is not only a novel way to evidence the uniform-staggered transition but also of fundamental interest for the study of collective modes in superconductors~\cite{Gassner2023}. However, the theory presented in  Ref.~\cite{Lee2022} does not explicitly account for gauge invariance or the Coulomb interaction, and not all the signatures of the BS mode in the electromagnetic (EM) response were evaluated. 

In this letter, we theoretically examine the EM response of a minimal model of a superconductor with a sublattice degree of freedom and a subdominant instability towards the staggered state, namely a bilayer with SOC. The EM response of bilayer superconductors has been extensively studied in the context of bilayer cuprates~\cite{WenChin1995,Forsthofer1996,Chaloupka2009thesis,Chaloupka2009,Sellati2023}, where the pairing potential is comparable to the band splitting and SOC is negligible. Here we are more interested in the limit where the pairing potentials are small compared to the band splitting and SOC is strong, which is relevant to CeRh$_2$As$_2$ and TMDs. To achieve a manifestly gauge-invariant theory, we treat the fluctuations into the staggered channel as short-wavelength phase fluctuations which can be naturally incorporated with external EM fields into a gauge-invariant action. We derive the EM response tensor in the long-wavelength limit and hence demonstrate the existence of an antisymmetric BS-like phase mode within the superconducting excitation gap. However, this phase mode couples directly to the Coulomb interaction, which likely pushes the mode energy into the quasiparticle continuum and thus impedes experimental observation. 

{\it Theoretical model.} We consider a two-dimensional square bilayer model with in-plane lattice constant $a$ and layer separation $\delta$. This is described by the Lagrangian 
\begin{eqnarray} \label{eq:lagrangian}
  L &=& \sum_{{\bf r},{\bf r}'} \left\{\mathbf{c}^\dagger_{\bf r} \left( \delta_{{\bf r},{\bf r'}} \hbar \partial_\tau + \mathcal{H}_0(\mathbf{p} +e\mathbf{A}) -e\delta_{{\bf r},{\bf r'}} {\phi} \right) \mathbf{c}_{\bf r'} \right\} \nonumber\\ 
  &&- 4 \sum_{\bf r} \sum_{\eta = A,B} g c_{\eta,{\bf r},\uparrow}^\dagger c_{\eta,{\bf r},\downarrow}^\dagger c_{\eta,{\bf r},\downarrow} c_{\eta,{\bf r},\uparrow},
\end{eqnarray}
where ${\bf c}_{\bf r} = (c_{A,{\bf r},\uparrow}, c_{A,{\bf r},\downarrow}, c_{B,{\bf r},\uparrow}, c_{B,{\bf r},\downarrow})^T$ is a spinor of fermion annihilation operators accounting for spin and layer ($A$, $B$) degrees of freedom, $\mathbf{p}$ is the momentum operator, $A^\mu = (\phi, \mathbf{A})$ is the EM four-potential of an external field, and $e$ is the fundamental charge.
The normal state Hamiltonian matrix $\mathcal{H}_{0}({\bf k})$ is constrained by the tetragonal symmetry of the bilayer and has the general form~\cite{Yoshida2012,Yoshida2014}
\begin{equation} \label{eq:normal_state}
    \mathcal{H}_{0}(\mathbf{k})= \epsilon_{00,\mathbf{k}} \eta_0 \sigma_0 + \epsilon_{x0,\mathbf{k}} \eta_x \sigma_0 + \epsilon_{zx,\mathbf{k}} \eta_z \sigma_x + \epsilon_{zy,\mathbf{k}} \eta_z \sigma_y,
\end{equation}
where the $\eta_i$ and $\sigma_i$ Pauli matrices encode the layer and spin degrees of freedom. Here we take
$\epsilon_{00,\mathbf{k}}=-2t[\cos(k_x a) + \cos(k_y a)] - \mu$ where $\mu$ is the chemical potential and $t$ is the hopping between nearest-neighbour sites in the same layer. $\epsilon_{zx,\mathbf{k}}=\alpha \sin(k_y a)$ and $\epsilon_{zy,\mathbf{k}}=-\alpha \sin(k_x a)$ describe Rashba SOC of strength $\alpha$ which originates from the locally-broken inversion symmetry on each layer and reverses sign between the two layers to preserve the global inversion symmetry. Finally, $\epsilon_{x0,\mathbf{k}}=t_\perp$ is the interlayer hopping between the $A$ and $B$ sites in each unit cell.
The Hamiltonian describes a two-band system with energies $\xi_{j=1,2,{\bf k}} = \epsilon_{00,{\bm k}} - (-1)^{j} \sqrt{\epsilon_{zx,\mathbf{k}}^2+\epsilon_{zy,\mathbf{k}}^2+\epsilon_{x0,\mathbf{k}}^2}$. 

The second term in Eq.~\ref{eq:lagrangian} is an attractive interaction which mediates on-site $s$-wave spin-singlet pairing. This generates a pairing instability in two channels: the even-parity uniform state $\hat{\Delta}_u = \Delta_u \eta_0i\sigma_y$, where the pair potentials are the same on both layers, and the odd-parity staggered state $\hat{\Delta}_s = \Delta_s \eta_zi\sigma_y$, where the sign of the potential reverses between the two layers. The uniform state pairs electrons in the same band and opens a gap $|\Delta_u|$ at the Fermi energy. In contrast, the staggered state also involves interband pairing, which reduces the gap at the Fermi energy below $|\Delta_s|$. SOC must be present for the staggered state to pair electrons in the same band; in the limit where the band splitting is much larger than the pairing potential, the gap at the Fermi surface is
given by $|\Delta_s|\sqrt{F_{\bf k}}$ where $F_{\bf k} = 4(\epsilon_{zx,{\bf k}}^2 + \epsilon_{zy,{\bf k}}^2)/(\xi_{1,{\bf k}}-\xi_{2,{\bf k}})^2\leq1$ is the 
superconducting fitness~\cite{Ramires2018}. Despite having the same pairing interaction, the generically smaller gap opened by the staggered state compared to the uniform state implies that the latter is the stable superconducting phase. 

The EM field is included within the minimal coupling scheme. Using the Peierls substitution, we expand the Hamiltonian up to second order
\begin{align} \label{eq:EMexpand}
&\sum_{\mathbf{r,r^\prime}} 
\mathbf{c}^\dagger_{\bf r}  \left(\mathcal{H}_0(\mathbf{p} +e\mathbf{A})  -e\delta_{{\bf r},{\bf r'}}\phi \right) \mathbf{c}_{\bf r^\prime} \nonumber\\
&\approx \sum_{\mathbf{r,r^\prime}} \mathbf{c}^\dagger_{\bf r} \mathcal{H}_0({\bf p}) \mathbf{c}_{\bf r^\prime} 
 + \sum_{\mathbf{r}} \mathcal{J}_{\mu}({\bf r})A^\mu({\bf r}) \nonumber \\
 &  \phantom{\approx \sum_{\mathbf{r,r^\prime}}}+ \frac{1}{2}\sum_{\mathbf{r}}e^2\mathcal{D}_{ij}({\bf r})A_i({\bf r})A_j({\bf r}) ,
\end{align} 
where $\mathcal{J}_\mu$ is the paramagnetic current and $\mathcal{D}_{ij}$ is the diagmagnetic Drude kernel.
Although the EM field varies continuously in space, in our tight-binding model it is sampled at the discrete lattice points. We accordingly define $A^\mu_{\eta,\mathbf{r_j}}$ as the EM four-potential at lattice site $\mathbf{r_j}$ in layer $\eta=A,B$. We then decompose the EM field in each layer in terms of symmetric and antisymmetric components $A^\mu_{+}$ and $A^\mu_{-}$~\cite{Liu2017}, respectively, as
\begin{eqnarray}
    A^\mu_{\pm} = \frac{1}{2} ( A^\mu_{A} \pm A^\mu_{B}) . \label{eq:EMdecomp}
\end{eqnarray}
The paramagnetic and diamagnetic terms in Eq.~\ref{eq:EMexpand} are then written in terms of the symmetric and antisymmetric EM field components
\begin{eqnarray}
    \mathcal{J}_{\mu} A^\mu &=& \sum_{a=\pm} \mathcal{J}_{a,\mu} A^\mu_a, \\
    \mathcal{D}_{ij} A_i A_j &=& \sum_{a,b=\pm} [\mathcal{D}_{ab}]_{ij} A_{a,i} A_{b,j} .
\end{eqnarray}
The paramagnetic current operators are most conveniently defined in momentum space ${\cal J}^\mu_{\pm}({\bf q}) = -e \sum_{\bf k}{\bf c}^\dagger_{\bf k+q} {J}^\mu_\pm({\bf k},{\bf q}){\bf c}_{\bf k}$ where the matrix elements are
\begin{eqnarray}
    {J}^\mu_+({\bf k},{\bf q}) &=& (\eta_0\sigma_0, \tfrac{1}{2}[{\bf V}_{\bf k}+{\bf V}_{\bf{k+q}}] + v_{x0} \eta_y\sigma_0 \hat{\mathbf{z}}) \label{eq:symPara}, \\
    {J}^\mu_-({\bf k},{\bf q}) &=& (\eta_z\sigma_0, \tfrac{1}{2}\eta_z[{\bf V}_{\bf k}+{\bf V}_{\bf{k+q}}] ) .
\end{eqnarray}
Here ${\bf V}_{\bf k} = \frac{1}{\hbar} \partial_\mathbf{k}\mathcal{H}_0$, 
where the derivative is with respect to the in-plane momenta, i.e. $\partial_\mathbf{k} \equiv \partial_{k_x}\hat{\mathbf{x}} + \partial_{k_y} \hat{\mathbf{y}}$, and $v_{x0} =t_\perp\delta / \hbar$ is the contribution of the interlayer hopping to the paramagnetic current. Treating the diamagnetic current similarly, the corresponding momentum-space matrix elements at ${\bf q}=0$ are 
\begin{eqnarray}
    \left [D_{++}\right]_{ij} &=& \frac{1}{\hbar^2} \left( \partial_{k_i k_j} \mathcal{H}_0 -\delta_{ij} \delta_{jz} t_\perp \delta^2 \eta_x\sigma_0 \right)
  , \label{eq:symDia}\\
    \left[D_{--}\right]_{ij} &=& \frac{1}{\hbar^2} \partial_{k_i k_j} \mathcal{H}_0, \\
    \left[D_{+-}\right]_{ij} &=& \left[D_{-+}\right]_{ij} = \frac{1}{\hbar^2}  \partial_{k_i k_j} \eta_z\mathcal{H}_0 . 
\end{eqnarray} 

{\it Effective action.} We decouple the interaction term in the pairing channels using the Hubbard-Stratanovich transformation and then integrate out the fermions to obtain an  effective action for the EM field and the pairing potentials alone
\begin{equation} \label{eq:eff_action}
    S[A,\Delta] = \text{Tr ln} \ G_{A \Delta} + \int d\tau \sum_{\bf r} \frac{(\bar{\Delta}_u \Delta_u + \bar{\Delta}_s \Delta_s)}{2g} ,
\end{equation}
where $G_{A\Delta}$ is the full fermion propagator which accounts for the coupling to the pairing fields, $\Delta_u$ and $\Delta_s$, and an external EM field. The trace is over the frequency, momentum, spin, and layer degrees of freedom of the fermions.

The aim of this work is to determine an action for the EM field only, i.e. 
\begin{align} \label{eq:gauge_inv_action}
    S_\text{EM} [A] = \sum_q \sum_{a,b=\pm} e^2 \Pi^{\mu\lambda}_{ab} (q) A_{a,\mu}(-q) A_{b,\lambda} (q) ,
\end{align} 
where $\Pi^{\mu\lambda}$ is the gauge-invariant EM response tensor and $q=(i\omega,{\bf q})$ accounts for the frequency and momentum of the fluctuating fields. Since gauge invariance is equivalent to requiring charge conservation, it is enlightening to consider charge conservation in a bilayer system.
As pointed out in Refs.~\cite{Chaloupka2009thesis,Chaloupka2009}, we must distinguish between processes which change the total charge in a unit cell, and processes which redistribute charge between the different layers in a unit cell without changing the total charge. The former processes are analogous to the usual currents in a monolayer system and give rise to the conservation law for the symmetric currents $q_\mu{\cal J}^\mu_+(q) = 0$. 
In contrast, the latter processes involve the antisymmetric currents ${\cal J}^{0,x,y}_-$ and the symmetric out-of-plane current ${\cal J}^z_+$, with the conservation law $q_\mu{\cal J}^{\mu}_-(q) + \frac{2i}{\delta}{\cal J}^z_+(q) = 0$. From the definition of the EM response tensor we have $\langle {\cal J}^\mu_a(q) \rangle = -e^2\Pi^{\mu\lambda}_{ab}A_{b,\lambda}(q)$; combining this with the charge conservation laws, we have the conditions on the EM response tensor
\begin{align}\label{eq:gaugeCondition}
    q_\mu \Pi^{\mu\lambda}_{+a} = 0, \quad q_\mu \Pi^{\mu\lambda}_{-a} + \frac{2 i}{\delta} \Pi^{z \lambda}_{+a} = 0,
\end{align}
where $a=\pm$. Note that the second condition includes a contribution from interlayer currents which does not vanish in the long-wavelength (i.e. ${\bf q}\rightarrow0$) limit.

As discussed above, the uniform state is realized at the saddle-point with pairing potential $\Delta_u = \Delta_0$ which we choose to be real. The attractive interaction in the subdominant staggered channel is then expected to generate a BS mode with energy within the excitation gap $2\Delta_0$.
In going beyond the mean-field solution, the literature on collective modes in superconductors~\cite{Allocca2019,Sun2020,Poniatowski2022} leads us to adopt the {\it Ans\"atze} for the pairing potentials in layers $A$ and $B$ in unit cell $\mathbf{r_j}$ 
\begin{align}
    \Delta_{A,\mathbf{r_j}} &= [\Delta_0 + \delta\Delta^r_u(\mathbf{r_j}) + \delta\Delta^r_s(\mathbf{r_j}) + i\delta\Delta^i_s(\mathbf{r_j})] e^{2i\theta_{\mathbf{r_j}}} \label{eq:DeltaA} ,\\
    \Delta_{B,\mathbf{r_j}} &= [\Delta_0 + \delta\Delta^r_u(\mathbf{r_j}) - \delta\Delta^r_s(\mathbf{r_j}) - i\delta\Delta^i_s(\mathbf{r_j})] e^{2i\theta_{\mathbf{r_j}}} \label{eq:DeltaB} ,
\end{align}
which includes fluctuations in the overall phase $\theta$, the amplitude in the uniform channel $\delta\Delta^r_u$, and both real and imaginary fluctuations in the staggered channel, $\delta\Delta^r_s$ and $\delta\Delta^i_s$, respectively. Indeed, this is the approach taken by Ref.~\cite{Lee2022} to study a similar system. In the appendix, we follow the method of Ref.~\cite{Lee2022} to construct the Gaussian action for the imaginary fluctuations in the staggered channel and verify that it gives rise to a subgap BS mode~\footnote{See the appendix for a summary of the method in Ref.~\cite{Lee2022}, explicit expressions for the EM kernel, and the inclusion of the Coulomb interaction.}. Integrating out these fluctuations we obtain an EM-only action, which although not obvious {\it a priori}, does satisfy the second gauge-invariance condition. Surprisingly, the theory is \emph{only} gauge invariant upon the inclusion of the coupling to the BS mode, which is not the case in other systems with subdominant pairing channels~\cite{Sun2020,Muller2021}.

To understand the critical role of fluctuations in the staggered channel, we recall the standard approach to construct the gauge-invariant form of the action by combining the EM field with the superconducting phase \cite{Paramekanti2000,Benfatto2004}. The {\it Ans\"atze} Eqs.~\ref{eq:DeltaA} and~\ref{eq:DeltaB} are therefore deficient, as the phase locking between the layers prevents us from formulating a manifestly gauge-invariant theory for all the components of the EM field.
This can be remedied by the modified {\it Ans\"atze}
\begin{align}
    \Delta_{A,\mathbf{r_j}} &= [\Delta_0 + \delta\Delta^r_u(\mathbf{r_j}) + \delta\Delta^r_s(\mathbf{r_j})] e^{2i\theta_{A,\mathbf{r_j}}} \label{eq:DeltaA1} ,\\
    \Delta_{B,\mathbf{r_j}} &= [\Delta_0 + \delta\Delta^r_u(\mathbf{r_j}) - \delta\Delta^r_s(\mathbf{r_j})] e^{2i\theta_{B,\mathbf{r_j}}} \label{eq:DeltaB1} ,
\end{align}
where we now allow the phase of the order parameter to vary between the two layers. Analogous to  our decomposition of the EM field, we express the phase in each layer in terms of symmetric and antisymmetric components
\begin{eqnarray}
    \theta_A = \theta_u + \theta_s,\quad \theta_B = \theta_u - \theta_s,
\end{eqnarray}
where the subscripts on the RHS identify $\theta_u$ and $\theta_s$ with the uniform and staggered pairing states, respectively. The imaginary fluctuations in the staggered channel are now accounted for by the antisymmetric phase fluctuations $\theta_s$. 

We remove phase fluctuations from the order parameter by performing the local gauge transformation 
\begin{equation} \label{eq:gauge}
    c_{\eta,\mathbf{r_j},\sigma} \rightarrow e^{i \theta_{\eta,\mathbf{r_j}}} c_{\eta,\mathbf{r_j},\sigma}.
\end{equation}
In the long-wavelength limit this modifies the symmetric and antisymmetric EM field components as
\begin{eqnarray}
   (e\phi_+, e\mathbf{A}_+) &\rightarrow& (e\phi_+ - i \hbar \partial_\tau\theta_u, e\mathbf{A}_+ + \hbar (\nabla \theta_u + \tfrac{2}{\delta} \theta_s\hat{\mathbf{z}})),\notag\\ \label{eq:gaugeEM+} \\
    (e\phi_-,e\mathbf{A}_-) &\rightarrow& (e\phi_- - i\hbar\partial_\tau\theta_s, e \mathbf{A}_- + \hbar \nabla \theta_s) \label{eq:gaugeEM-},
\end{eqnarray} 
where 
$\nabla \equiv \partial_x\hat{\mathbf{x}} + \partial_y\hat{\mathbf{y}}$. 
Focusing on antisymmetric phase fluctuations, we see that these
couple to the antisymmetric scalar and in-plane vector potentials, as well as the  $z$-component of the symmetric vector potential. We isolate these in the mixed EM four-potential
\begin{equation} \label{eq:mixed}
    \Tilde{A}^\mu \equiv (\phi_-, (\mathbf{A}_-)_x, (\mathbf{A}_-)_y, (\mathbf{A}_+)_z).
\end{equation}
Under the gauge transformation Eq.~\ref{eq:gauge} this four-vector transforms as 
$e\Tilde{A}^\mu \rightarrow e\Tilde{A}^\mu - \hbar\Tilde{\partial}^\mu \theta_s$,
where we define $\Tilde{\partial}_\mu \equiv (i\partial_\tau, \nabla + \tfrac{2}{\delta}\hat{\mathbf{z}})$.

{\it EM response tensor.} We decompose the effective action
$S[A,\Delta]=S_{\text{mf}}+S_{\text{fluc}}$ where $S_{\text{mf}}$ corresponds to evaluating the trace with respect to the static mean-field solution for the order parameters, and the fluctuation action $S_{\text{fluc}}$ is the difference between the mean-field and full actions. Since we are primarily interested in the signatures of the odd-parity pairing channel, we only consider terms in the action which are coupled to the fluctuating phase $\theta_s$. In our bilayer model, fluctuations in the uniform channel decouple from $\theta_s$ and can be neglected. Consistent with previous work~\cite{Sun2020,Allocca2019}, real fluctuations in the staggered channel $\delta\Delta^r_s$ only introduce a small correction to the low-energy response through their coupling to the EM field and are also neglected. 

Expanding to second order in the EM field and phase fluctuations we obtain the Gaussian action
\begin{equation} \label{eq:EMaction}
    S [\theta_s,\Tilde{A}] = \frac{1}{2} \sum_q K^{\mu\lambda}(q) (e\Tilde{A}_\mu - \hbar\Tilde{\partial}_\mu \theta_s)_{-q} (e\Tilde{A}_\lambda - \hbar\Tilde{\partial}_\lambda \theta_s)_{q} .
\end{equation}  
Explicit expressions for the components of the EM kernel $K^{\mu\nu}$ are given in the appendix~\cite{Note1}. To obtain the EM-only action we integrate out the phase $\theta_s$ to obtain
\begin{align} \label{eq:gaugeAction}
      S_\text{EM} [\Tilde{A}] =\frac{1}{2} \sum_q e^2 \Pi^{\mu\lambda}(q) \Tilde{A}_\mu (-q)\Tilde{A}_\lambda (q) ,
\end{align} 
where the EM response tensor is given by 
\begin{align} \label{eq:Pi}
    \Pi^{\mu\lambda} (q) = K^{\mu\lambda}(q)  - \frac{ \Tilde{q}_\alpha  \Tilde{q}^\ast_\beta  K^{\alpha\lambda}(q)K^{\mu\beta}(q)}{ \Tilde{q}_a \Tilde{q}^\ast_b K^{ab}(q)}.
\end{align}
Here we define $\Tilde{q}^\ast_\mu \equiv (i\omega, -i\mathbf{q} + \frac{2}{\delta}\hat{\mathbf{z}})$ and $\Tilde{q}_\mu \equiv (-i\omega, i\mathbf{q} + \frac{2}{\delta}\hat{\mathbf{z}})$. It is straightforward to verify that the EM tensor satisfies the second gauge-invariance condition in Eq.~\ref{eq:gaugeCondition}. Upon analytic continuation to real frequency, $\Pi^{\mu\lambda}$ is directly related to observable quantities of the system such as the optical conductivity $\sigma_{zz} = \frac{i e^2}{\omega}\Pi^{zz}$. 

\begin{figure}[t]
    \centering
    \begin{overpic}[width=\columnwidth]{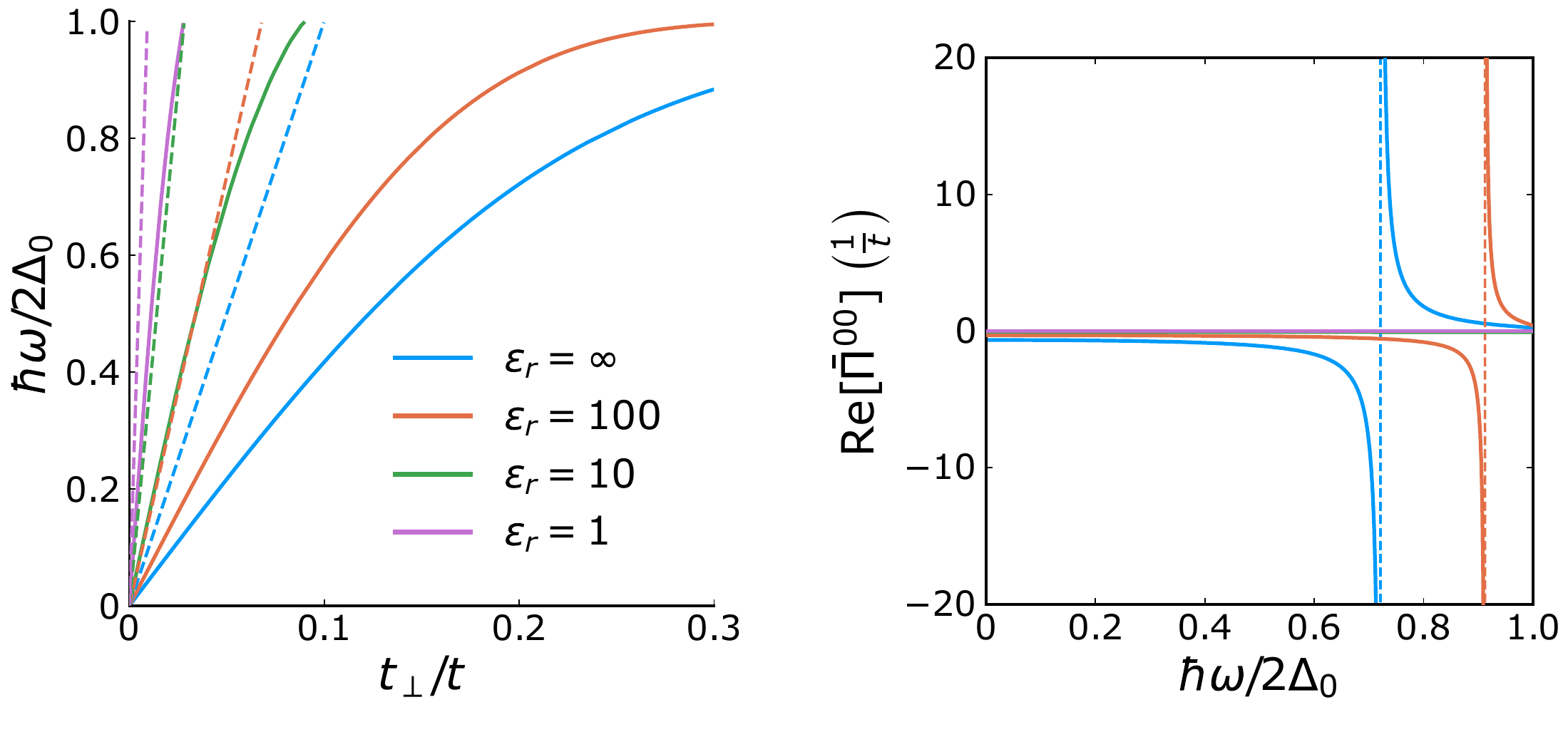}
    \put(2, 2){(a)}
    \put(54, 2){(b)}
  \end{overpic}
    \caption{{\bf (a)} Energy of the antisymmetric phase mode as a function of the interlayer hopping $t_\perp$ at varying relative permittivities. Dashed lines are calculated in the limit of vanishing SOC $\alpha=0$, and solid lines correspond to nonzero SOC $\alpha = 0.3t$. The $\epsilon_{r}=\infty$ curve corresponds to mode energy in the absence of the Coulomb interaction, i.e. the pole of Eq.~\ref{eq:antiDensity}. {\bf (b)} Plot of $\bar{\Pi}^{00}$ at varying permittivities for $\alpha=0.3t$ and $t_\perp=0.2t$. For these plots we set $a=0.5$~nm, $\delta=2a$, $t=0.1$~eV and $\Delta_0=0.1t$. Simulated on an $N\times N$ square lattice with $N=800$.}
    \label{fig:genFreq}
\end{figure}

We now focus on the antisymmetric density-density correlator $\Pi^{00}$ at $\mathbf{q}=0$
\begin{align} \label{eq:antiDensity}
    \Pi^{00} = \frac{\frac{4}{\delta^2}(K^{00}K^{zz}-K^{0z}K^{z0})}{\omega^2K^{00}+\frac{2i\omega}{\delta} (K^{z0} - K^{0z})+\frac{4}{\delta^2}K^{zz}} .
\end{align}
By the second gauge-invariance condition \ignore{, at $\mathbf{q}=0$,} the other relevant components of the EM response tensor are given by $\Pi^{0z} = \frac{\delta i \omega}{2} \Pi^{00}$ and $\Pi^{zz} = \frac{\delta i \omega}{2} \Pi^{z0}$.
We find a pole in the EM tensor within the excitation gap, corresponding to an antisymmetric phase mode. 
The mode energy as a function of interlayer hopping $t_\perp$ for fixed SOC $\alpha$ is given by the $\epsilon_r=\infty$ curve shown in Fig.~\ref{fig:genFreq}. The energy of the mode is independent of the distance $\delta$ between the layers. The mode energy vanishes in the limit of decoupled layers (i.e. $t_\perp\rightarrow 0$) where the system possesses global $U(1)$ gauge symmetry with respect to the phase in each layer independently. At nonzero interlayer hopping the mode energy depends crucially upon the SOC: in the absence of SOC it only occurs at subgap energies for sufficiently small interlayer hopping~\cite{WenChin1995,Forsthofer1996}, but when SOC is present it lies within the gap for all interlayer hopping strengths. The dramatic effect of the SOC reflects its role in stabilizing a weak-coupling instability in the staggered channel. As shown in the appendix~\cite{Note1}, this phase mode displays the same dependence on model parameters as the BS mode introduced through imaginary fluctuations in the staggered channel $\delta\Delta^i_s$ found in Ref.~\cite{Lee2022} using the order parameter {\it Ans\"atze} in Eqs.~\ref{eq:DeltaA}~and~\ref{eq:DeltaB}. 

{\it Coulomb interaction.} Phase fluctuations in a superconductor couple directly to density fluctuations, which in a charged system are subject to the Coulomb interaction.  
In the bilayer, the long-range Coulomb interaction can be decomposed in terms of symmetric and antisymmetric components
\begin{align}
     V_{\pm,\mathbf{q}} &= \frac{1}{a^2}\frac{\pi e^2}{\epsilon_b}\frac{(1\pm e^{-|\mathbf{q}|\delta})}{|\mathbf{q}|} \label{eq:coulomb},
\end{align}
where $V_{+}$ and $V_{-}$ couple to symmetric and antisymmetric density fluctuations, respectively, and $\epsilon_b = 4\pi\epsilon_0\epsilon_r$ is the dielectric constant of the ionic background which depends on the relative permittivity $\epsilon_r$. As shown in the supplemental material~\cite{Note1}, the antisymmetric Coulomb interaction modifies the density-density response Eq.~\ref{eq:antiDensity} as
\begin{align}
    \Pi^{00} \rightarrow \frac{\Pi^{00}}{1-V_- \Pi^{00}} \equiv \bar{\Pi}^{00} .
\end{align}
The energy of the antisymmetric phase mode is given by the pole of $\bar{\Pi}^{00}$ and now depends on the strength of the Coulomb interaction; we plot the mode energy for differing relative permittivities in Fig.~\ref{fig:genFreq}(a). The Coulomb interaction increases the mode energy so that we only find a subgap excitation for sufficiently weak interlayer coupling; for realistic parameter choices $\delta \geq 2a$, $\alpha \lesssim t_\perp$ and $\epsilon_r \lesssim 10$, the mode energy lies outside of the superconducting gap.
Accordingly, the EM response functions become featureless at subgap energies as shown in Fig.~\ref{fig:genFreq}(b).

To understand this result, we consider the limit when the two layers are decoupled, i.e. $t_\perp = 0$. The independent variation of the phase in each layer gives rise to two degenerate Anderson-Bogoliubov-Goldstone (ABG) phase modes. Introducing interlayer hopping hybridizes these modes into symmetric and antisymmetric combinations, where the former is the ABG mode of the bilayer and the latter is the mode found above. The evolution of the antisymmetric phase mode from an ABG mode makes it distinct from other BS modes~\cite{Bardasis1961,Maiti2016,Sun2020,Muller2021}, and also explains its sensitivity to the Coulomb interaction. Specifically, the symmetric and antisymmetric phase modes are converted to symmetric and antisymmetric plasmons~\cite{WenChin1995,Forsthofer1996,DasSarmabilayer}, respectively. In the presence of interlayer tunneling the antisymmetric plasmon acquires a gap~\cite{DasSarmabilayer}, and with increasing $t_\perp$ is pushed out of the superconducting excitation gap, taking the phase mode with it, as seen in Fig.~\ref{fig:genFreq}(a). Although the bare mode energy is predicted to significantly soften as we approach the critical Zeeman splitting $\sim\Delta_0$~\cite{Lee2022}, this should not affect the energy $\sim t_\perp\gg \Delta_0$ of the antisymmetric plasmon. We hence do not expect to see the mode to appear at subgap energies close to the even-odd transition.
In this regard, the fate of the antisymmetric phase mode is similar to that of the ABG mode in a three-dimensional superconductor. 

{\it Conclusions.} We have constructed a gauge-invariant theory of the electromagnetic response of a superconducting bilayer with strong spin-orbit coupling. To fully account for charge conservation in the bilayer~\cite{Chaloupka2009,Chaloupka2009thesis}, we treat fluctuations in the staggered pairing channel as short-wavelength phase fluctuations, which directly couple to the EM field in a way that preserves gauge invariance. This gives rise to a low-lying pole in the EM response tensor corresponding to an antisymmetric phase mode, which can be considered as a BS mode arising from fluctuations into the staggered state. However, the origin of this mode from phase fluctuations converts it into a plasmon,  preventing its experimental observation. In this work, we have focused on a superconducting bilayer as a minimal model of a superconductor with a sublattice degree of freedom. Our conclusions likely also apply to systems with more complicated geometries, e.g. the honeycomb lattice or the nonsymmorphic structure of CeRh$_2$As$_2$. A more promising setting in which to search for this mode is neutral cold atomic gases in an optical lattice mimicking a spin-orbit-coupled bilayer~\cite{Wang2017}.

{\it Acknowledgements.} The authors acknowledge stimulating discussions with O. Sushkov, and conversations with D. F. Agterberg, J. Link, and C. Timm. We are grateful to C. Lee and S. B. Chung for sharing their preprint~\cite{Lee2022} before submission to the arXiv and for discussions of their work. PMRB was supported by the Marsden Fund Council from Government funding, managed by Royal Society Te Ap\={a}rangi, Contract No. UOO2222.

\bibliography{refs}

\pagebreak
\onecolumngrid
\appendix

\section{BS mode and gauge invariance} \label{app:BSmode}
\begin{figure}
    \centering
    \begin{overpic}[width=0.8\columnwidth]{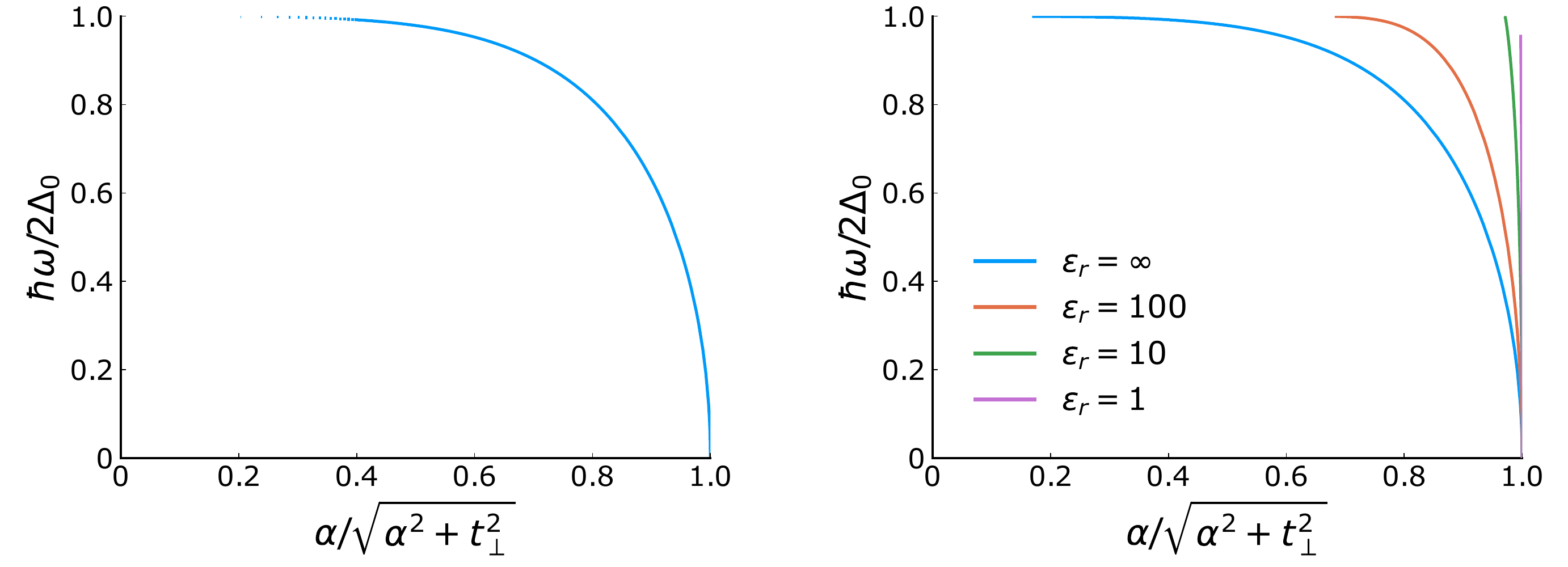}
    \put(2, 2){(a)}
    \put(54, 2){(b)}
  \end{overpic}
     \caption{{\bf (a)} Energy of the BS mode and {\bf (b)} antisymmetric phase mode as a function of the SOC strength $\alpha$ relative to the interlayer hopping $t_\perp$. We vary both parameters such that $\sqrt{\alpha^2+t_\perp^2} = 0.5t$ is constant. The antisymmetric mode energy includes renormalization due to the Coloumb interaction and is plotted at varying permittivities with $\epsilon_r = \infty$ corresponding to the limit in which the Coulomb interaction is vanishing. All calculations are simulated on an $N\times N$ square lattice with $N=800$.} \label{fig:BSFreq}
\end{figure}

In this section, we discuss the {\it Ans\"atze} for the fluctuation order parameters $\Delta_{A,\mathbf{r_j}}$ and $\Delta_{B,\mathbf{r_j}}$ shown in Eqs.~15 and~16 in the main text. The overall phase in each unit cell $\theta_{\mathbf{r_j}}$ can be combined with the EM field by performing the gauge transformation
\begin{equation} \label{eq:gauge1}
    c_{\eta,\mathbf{r_j},\sigma} \rightarrow e^{i \theta_{\mathbf{r_j}}} c_{\eta,\mathbf{r_j},\sigma},
\end{equation}
where, in contrast to Eq. 20, the transformation is independent of the layer index. This transformation does not modify $\Tilde{A}_\mu$. We now focus on the imaginary fluctuations in the staggered channel $\Delta^i_s$ which give rise to a low-lying BS mode. The Gaussian action describing these fluctuations is
\begin{align}
    S_\text{BS} = \frac{1}{2} \sum_q G^{-1}_\text{BS} (q) \delta\Delta^i_s (-q) \delta\Delta^i_s (q), 
\end{align}
where $G^{-1}_\text{BS}$ is inverse propagator for the BS mode 
\begin{align} \label{eq:BSprop}
    G^{-1}_\text{BS} (q) = \frac{1}{g} + \chi_\text{BS} (q) .
\end{align}
The relevant two-particle correlator is defined by
\begin{align}
    \chi_\text{BS} (q) =& \frac{1}{2} \sideset{}{'}\sum_k \text{Tr} \left[ G(k) (\tau_x\eta_z\sigma_y) G(k+q) (\tau_x\eta_z\sigma_y) \right],
\end{align}
where $k=(i\omega_n,\mathbf{k})$, $G$ is the fermionic propagator evaluated at the mean-field solution and the $\tau_i$ matrices encode the particle-hole degree of freedom. We adopt the notation $\sum^\prime_{k} \equiv (k_BT/N)\sum_{i\omega_n}\sum_\mathbf{k}$ to account for the normalised momentum and fermionic Matsubara sums.  We analytically continue the correlator to real frequencies with the replacement $i\Omega\rightarrow \omega +i0^+$. 
In the $\mathbf{q}\rightarrow 0$, $T\rightarrow 0$ limit, the real-frequency correlator is
\begin{align} \label{eq:BScor}
    \chi_\text{BS} (\omega) = -\frac{1}{N} \sum_{\mathbf{k}} \Bigg\{ \sum_{j=1,2} F_\mathbf{k} \frac{4E_{j,\mathbf{k}}}{4E_{j,\mathbf{k}}^2-(\hbar \omega)^2} + (1-F_\mathbf{k}) \frac{2(E_{1,\mathbf{k}}+E_{2,\mathbf{k}})(E_{1,\mathbf{k}}E_{2,\mathbf{k}}+\Delta_0^2+\xi_{1,\mathbf{k}}\xi_{2,\mathbf{k}})}{ E_{1,\mathbf{k}}E_{2,\mathbf{k}}((E_{1,\mathbf{k}}+E_{2,\mathbf{k}})^2-(\hbar \omega)^2)} \Bigg\}. 
\end{align} 
Here $E_{1(2),{\bm k}}=\sqrt{\xi_{1(2),{\bf k}}^2 + \Delta_0^2}$ is the quasiparticle dispersion of the $1(2)$-band at the mean-field solution. The energy of the BS mode corresponding to the staggered pairing channel is determined by solving $\chi_{\text{BS}}(q)=-1/g$ which corresponds to a pole in the BS propagator $G_\text{BS}$. The BS frequency $\omega_\text{BS}$ is plotted as a function of SOC strength in Fig.~\ref{fig:BSFreq}(a). The energy of the antisymmetric phase mode corresponding to the pole of Eq.~29 in the main text is plotted in Fig.~\ref{fig:BSFreq}(b) for comparison. In the absence of the Coulomb interaction, i.e. $\epsilon_r = \infty$, we see that the dependence of the antisymmetric phase mode on the model parameters is identical to that of the BS mode. This illustrates that the two modes are indeed equivalent.

Interestingly, as discussed in Ref.~[19], the BS mode couples to the EM field in the linear regime. 
At the Gaussian level, the coupling is described by the action 
\begin{align}
    S_\text{c} = \sum_{q,a} C_{a,\mu} (q) e A^\mu_a(-q) \delta\Delta^i_s(q),
\end{align}
where the coupling coefficients are given by
\begin{align} \label{eq:BScoupling}
    C^\mu_a(q) = -\frac{1}{2}\sideset{}{'}\sum_k \text{Tr} \left[ G(k) J^\mu_a G(k+q) \left(\tau_x\eta_z\sigma_y\right) \right].
\end{align}
In the $\mathbf{q}\rightarrow 0$ limit, we find two nonzero coupling elements
\begin{align} 
    C^z_+(\omega) &= \frac{ \Delta_0}{N} \sum_\mathbf{k}  v_{x0} \frac{\epsilon_{x0,\mathbf{k}}}{ \sqrt{\epsilon_{zx,\mathbf{k}}^2+\epsilon_{zy,\mathbf{k}}^2+\epsilon_{x0,\mathbf{k}}^2}}  \frac{2 (E_{1,\mathbf{k}}+E_{2,\mathbf{k}})(\xi_{1,\mathbf{k}}-\xi_{2,\mathbf{k}})}{E_{1,\mathbf{k}}E_{2,\mathbf{k}} \left( (E_{1,\mathbf{k}}+E_{2,\mathbf{k}})^2 - (\hbar \omega)^2 \right)} ,\label{eq:couplingz}\\
    C^0_-(\omega) &=  -\frac{2i\Delta_0
    \hbar\omega}{N}  \sum_\mathbf{k}\Bigg\{ \sum_{j=1,2} F_\mathbf{k}\frac{1}{E_{j,\mathbf{k}}(4E_{j,\mathbf{k}}^2-(\hbar \omega)^2)} + (1-F_\mathbf{k}) \frac{E_{1,\mathbf{k}}+E_{2,\mathbf{k}}}{ E_{1,\mathbf{k}}E_{2,\mathbf{k}}((E_{1,\mathbf{k}}+E_{2,\mathbf{k}})^2-(\hbar \omega)^2)} \Bigg\} \label{eq:coupling0}.
\end{align}
The coupling of the BS mode to the $z$-component of the symmetric vector potential Eq.~\ref{eq:couplingz} was reported in Ref.~[19], however, the coupling to the antisymmetric scalar potential Eq.~\ref{eq:coupling0} was not fully addressed.
Neglecting phase fluctuations, the action for the EM field is given by
\begin{align}
    S[A] = \frac{1}{2} \sum_{q} \sum_{a,b=\pm} e^2 K^{\mu\lambda}_{ab} (q) A_{a,\mu} (-q) A_{b,\lambda} (q).
\end{align} 
Integrating out the imaginary odd-parity fluctuations modifies the EM kernel as
\begin{align}
    K^{\mu\lambda}_{ab} (q) \rightarrow K^{\mu\lambda}_{ab} (q) - G_\text{BS} (q) C^\mu_a (-q) C^\lambda_b (q) . 
\end{align}
Notice that at $\mathbf{q} = 0$, the BS mode couples only to the components of the EM field accounted for by the mixed four potential $\Tilde{A}_\mu$ Eq.~23. Thus, all relevant components of the EM kernel $K^{\mu\lambda}$ in Eq.~24 are modified by the BS mode. Surprisingly, the inclusion of this correction, together with properly accounting for overall phase fluctuations, is sufficient to produce an EM response tensor which numerically satisfies the gauge-invariance conditions Eq.~14. Note that this approach only produces a gauge-invariant EM response tensor when the coupling to the EM field is fully accounted for, i.e. you must incorporate both Eq.~\ref{eq:couplingz} and Eq.~\ref{eq:coupling0} into your theory. The recovery of gauge invariance in this approach is rather opaque. 
The introduction of a staggered phase, as in Eqs.~17 and~18, allows us to explicitly account for gauge invariance and more easily interpret the sensitivity of the mode to the Coulomb interaction. 

\section{EM kernel correlators} \label{app:EMkernel}
Here we present the correlation functions relevant to the EM action Eq.~24. 
At $\mathbf{q}=0$, the only relevant contributions to the action are
\begin{eqnarray}
    K^{00}(i\Omega) &=& \chi_{\rho\rho}^{--} (i\Omega), \\
    K^{0z}(i\Omega) &=& -K^{z0}(i\Omega) = \chi_{\rho j_z}^{-+} (i\Omega), \\
    K^{zz}(i\Omega) &=& \chi_{j_z j_z}^{++} (i\Omega) + \langle \Tilde{\cal D}_{zz} \rangle ,
\end{eqnarray}
where in each case the superscripts on the RHS correspond to the symmetric and antisymmetric components of the EM field. The relevant correlators are 
\begin{align}
    \chi_{\rho\rho}^{--} (i\Omega) &= \frac{1}{2}\sideset{}{'}\sum_{k} \text{Tr} \left[ G(k)\left(\tau_z J^0_-\right) G (k+q) \left(\tau_z {J}^0_-\right) \right], \\
    \chi_{\rho j_z}^{-+} (i\Omega) &= \frac{1}{2}\sideset{}{'}\sum_{k} \text{Tr} \left[ G(k)\left(\tau_z {J}^0_-\right)  G (k+q) \left(\tau_0 {J}^z_+\right) \right], \\
    \chi_{j_z j_z}^{++} (i\Omega) &= \frac{1}{2}\sideset{}{'}\sum_{k} \text{Tr} \left[ G(k)\left(\tau_0 {J}^z_+\right)  G (k+q) \left(\tau_0 {J}^z_+\right) \right], \\
    \langle \Tilde{\cal D}_{zz} \rangle &= \frac{1}{2}\sideset{}{'}\sum_k \text{Tr} \left[ G (k)\left(\tau_z { [{D}_{++}]_{zz}}\right) \right].
\end{align}
At $\mathbf{q}=0,$ in the $T\rightarrow 0$ limit, the real-frequency correlators are  
\begin{align}
     \chi_{\rho\rho}^{--} (\omega) &= -\frac{1}{N} \sum_\mathbf{k} \Bigg\{ \sum_{j=1,2} F_{\mathbf{k}} \frac{4\Delta_0^2}{E_{j,\mathbf{k}} (4 E_{j,\mathbf{k}}^2 - (\hbar \omega)^2)} + (1-F_{\mathbf{k}}) \frac{2(E_{1,\mathbf{k}}+E_{2,\mathbf{k}})(E_{1,\mathbf{k}}E_{2,\mathbf{k}}+\Delta_0^2-\xi_{1,\mathbf{k}}\xi_{2,\mathbf{k}})}{ E_{1,\mathbf{k}}E_{2,\mathbf{k}}((E_{1,\mathbf{k}}+E_{2,\mathbf{k}})^2-(\hbar \omega)^2)} \Bigg\} \label{eq:antiDensity1},\\
      \chi^{-+}_{\rho j_z}(\omega) &= - \frac{1}{N} \sum_{\mathbf{k}} v_{x0} \frac{\epsilon_{x0,\mathbf{k}}}{ \sqrt{\epsilon_{zx,\mathbf{k}}^2+\epsilon_{zy,\mathbf{k}}^2+\epsilon_{x0,\mathbf{k}}^2}} \frac{2 i (E_{1,\mathbf{k}} \xi_{2,\mathbf{k}} - E_{2,\mathbf{k}} \xi_{1,\mathbf{k}}) \hbar\omega}{E_{1,\mathbf{k}}E_{2,\mathbf{k}} \left( (E_{1,\mathbf{k}}+E_{2,\mathbf{k}})^2 - (\hbar \omega)^2 \right)}, \\
    \chi^{++}_{j_z j_z}(\omega) &= -\frac{1}{N} \sum_{\mathbf{k}} v_{x0}^2 \frac{2(E_{1,\mathbf{k}}+E_{2,\mathbf{k}})(E_{1,\mathbf{k}}E_{2,\mathbf{k}}-\Delta_0^2-\xi_{1,\mathbf{k}} \xi_{2,\mathbf{k}})}{E_{1,\mathbf{k}}E_{2,\mathbf{k}} \left( (E_{1,\mathbf{k}}+E_{2,\mathbf{k}})^2 - (\hbar \omega)^2 \right)}, \\
     \langle \Tilde{\cal D}_{zz} \rangle &= \frac{1}{\hbar^2 N} \sum_{\mathbf{k}} \sum_{j=1,2} (-1)^j t_\perp \delta^2 \frac{\epsilon_{x0,\mathbf{k}}}{{ \sqrt{\epsilon_{zx,\mathbf{k}}^2+\epsilon_{zy,\mathbf{k}}^2+\epsilon_{x0,\mathbf{k}}^2}}}  \frac{\xi_{j,\mathbf{k}}}{E_{j,\mathbf{k}}} .
\end{align} 

\section{Including the Coulomb repulsion} \label{app:Coulomb}

The Coulomb interaction in the bilayer is written
\begin{equation}
H_{\text{Coulomb}}= \frac{1}{2}\sum_{{\bf r}_j,{\bf r}_{j'}}\left\{V_{\text{intra}}({\bf r}_j-{\bf
  r}_{j'})\left(n_{t,{\bf r}_j}n_{t,{\bf r}_{j'}} + n_{b,{\bf
      r}_j}n_{b,{\bf r}_{j'}}\right) + V_{\text{inter}}({\bf r}_j-{\bf
  r}_{j'})\left(n_{t,{\bf r}_j}n_{b,{\bf r}_{j'}} + n_{b,{\bf
      r}_j}n_{t,{\bf r}_{j'}}\right)\right\}
\end{equation}
where the intralayer and interlayer potentials are given by
\begin{equation}
V_{\text{intra}}({\bf r}_j-{\bf
  r}_{j'}) = \frac{e^2}{4\pi\epsilon_0}\frac{1}{|{\bf r}_j-{\bf
  r}_{j'}|}\, \qquad V_{\text{inter}}({\bf r}_j-{\bf
  r}_{j'}) = \frac{e^2}{4\pi\epsilon_0}\frac{1}{\sqrt{|{\bf r}_j-{\bf
  r}_{j'}|^2 + \delta^2}}
\end{equation}
and ${\bf r}_j$, ${\bf r}_{j'}$ are in-plane position vectors.
Converting to momentum space we have
\begin{eqnarray}
H_{\text{Coulomb}} & = & \frac{1}{2N}\sum_{{\bf
    q}}\left\{V_{\text{intra}}({\bf q})\left(n_{t,-{\bf q}}n_{t,{\bf q}} + n_{b,-{\bf
      q}}n_{b,{\bf q}}\right) + V_{\text{inter}}({\bf q})\left(n_{t,-{\bf q}}n_{b,{\bf q}} + n_{b,-{\bf
      q}}n_{t,{\bf q}}\right)\right\} \notag\\
& = & \frac{1}{2N}\sum_{{\bf
    q}}\left\{V_{+}({\bf q})n_{+,-{\bf q}}n_{+,{\bf q}}+ V_{-}({\bf q})n_{-,-{\bf q}}n_{-,{\bf q}}\right\}
\end{eqnarray}
where $n_{\pm,{\bf q}} = n_{t,{\bf q}}\pm n_{b,{\bf q}}$ and $V_{\pm}({\bf q}) = \frac{1}{2}\left(V_{\text{intra}}({\bf q}) \pm
V_{\text{inter}}({\bf q})\right)$, which in the long-wavelength limit
(i.e. $|{\bf q}|a\ll 1$) are given by
\begin{equation}
V_{\pm}({\bf q}) = \frac{\pi e^2}{a^2\epsilon_b}\frac{\left(1\pm
    e^{-|{\bf q}|\delta}\right)}{|{\bf q}|} \,.
\end{equation}
where $\epsilon_b = 4\pi\epsilon_0\epsilon_r$. 

We decouple this by introducing the bosonic field $\rho_{+,{\bf q}}$
and $\rho_{-,{\bf q}}$, corresponding to symmetric and antisymmetric
density fluctuations, respectively. We accordingly perform the
Hubbard-Stratanovich decomposition
\begin{align}
\int D[\bar{c},c]&\exp\left(-\int^\beta_0 d\tau
  H_{\text{Coulomb}}\right) \notag \\
= & \int D[\bar{c},c,\rho_{+},\rho_{-}] \exp\left(-\int^\beta_0 d\tau \frac{1}{2N}\sum_{{\bf
    q}}\sum_{\nu=\pm}\left\{V_{\nu}({\bf q})n_{\nu,-{\bf
    q}}n_{\nu,{\bf q}} - \frac{\rho_{\nu,-{\bf
    q}}\rho_{\nu,{\bf q}}}{V_{\nu}({\bf q})}\right\}\right)\notag \\
= & \int D[\bar{c},c,\rho_{+},\rho_{-}]\exp\left(-\int^\beta_0 d\tau \frac{1}{2N}\sum_{{\bf
    q}}\sum_{\nu=\pm}\left\{-\rho_{\nu,{\bf q}}n_{\nu,-{\bf q}} - \rho_{\nu,-{\bf q}}n_{\nu,{\bf q}} - \frac{\rho_{\nu,-{\bf
    q}}\rho_{\nu,{\bf q}}}{V_{\nu}({\bf q})}\right\}\right)\,.
\end{align}
Note that $\rho_{+}$ and $\rho_{-}$ couple to the fermions in exactly
the same way as the scalar potentials $e\phi_{+}$ and $e\phi_{-}$,
respectively. 
We can hence modify our effective action to include the field $\rho_{-}$ as
\begin{equation}
S(e\tilde{A}_\mu -\hbar\tilde{\partial}_\mu\theta_s ) \rightarrow
S(e\tilde{A}_\mu -\hbar\tilde{\partial}_\mu\theta_s +
\rho_{-}\delta_{\mu,0}) - \sum_{q}\frac{1}{2V_{-}({\bf q})}\rho_{-,-{\bf q}}\rho_{-,{\bf q}}
\end{equation}
Upon integrating out the phase degree of freedom we thus have the
action
\begin{equation}
\frac{1}{2}\sum_{q}\left\{\Pi^{\mu\lambda}(q)(e\tilde{A}_\mu +
\rho_{-}\delta_{\mu,0})_{-q}(e\tilde{A}_\lambda +
\rho_{-}\delta_{\lambda,0})_{+q} - \frac{1}{V_{-}({\bf q})}\rho_{-,-q}\rho_{-,q}\right\}
\end{equation}
Performing the transformation $\rho_{-,q} \rightarrow
\rho_{-,q} + \left(\Pi^{00}(q)- \frac{1}{V_{-}({\bf
      q})}\right)^{-1}\Pi^{0\mu}e\tilde{A}_{\mu,q}$ we obtain
\begin{equation}
\frac{1}{2}\sum_{q}\left\{\left(\Pi^{00}(q)- \frac{1}{V_{-}({\bf
      q})}-\right)\rho_{-,-q}\rho_{-,q} + \left[\Pi^{\mu\lambda}(q) -
  \Pi^{\mu 0}(q)\left(\Pi^{00}(q)- \frac{1}{V_{-}({\bf
      q})}\right)^{-1}\Pi^{0\lambda}(q) \right]e^2\tilde{A}_{\mu,-q}\tilde{A}_{\lambda,q}\right\}
\end{equation}
We can safely drop the $\rho_{-,q}$ term to find the EM-only action
including the effect of the Coulomb interaction
\begin{equation}
\frac{1}{2}\sum_{q}\left[\Pi^{\mu\lambda}(q) -
  \Pi^{\mu 0}(q)\left(\Pi^{00}(q)- \frac{1}{V_{-}({\bf
      q})}\right)^{-1}\Pi^{0\lambda}(q) \right]e^2\tilde{A}_{\mu,-q}\tilde{A}_{\lambda,q}
\end{equation}
In particular, the renormalized density-density correlator is
\begin{equation}
\overline{\Pi}^{00}(q) = \Pi^{00}(q) -
  \Pi^{00}(q)\left(\Pi^{00}(q)- \frac{1}{V_{-}({\bf
      q})}\right)^{-1}\Pi^{00}(q) = \frac{\Pi^{00}(q)}{1 - V_{-}(q)\Pi^{00}(q)}
\end{equation}
which is the usual RPA result. The
renormalized correlators as equivalent to the Feynmann diagrams shown
in Fig. 45 of Ref. [32]. The poles of the renormalized
correlators are given by zeros of $1 - V_{-}(q)\Pi^{00}(q)$, which
reflects the conversion of the phase mode into an antisymmetric plasmon. The energy of the renormalized antisymmetric phase mode is plotted as a function of SOC strength at varying permittivities in Fig.~\ref{fig:BSFreq}(b).

\end{document}